\documentclass[final,5p,times,twocolumn]{elsarticle}
\usepackage{amssymb}
\usepackage{lipsum}
\usepackage{natbib}
\usepackage[colorlinks,linkcolor=blue]{hyperref}

\begin{document}

\begin{frontmatter}

\title{Four-Charm-Quark Matter from the CMS Collaboration as a Witness of the Development of High-Precision Hadron Spectroscopy}

\author[a1,a2,a3,a4,a5]{Xiang Liu\corref{cor1}}
\ead{xiangliu@lzu.edu.cn}

\cortext[cor1]{Corresponding author}

\affiliation[a1]{School of Physical Science and Technology, Lanzhou University, Lanzhou 730000, China}
\affiliation[a2]{Lanzhou Center for Theoretical Physics, Key Laboratory of Theoretical Physics of Gansu Province, Lanzhou University, Lanzhou 730000, China}
\affiliation[a3]{Key Laboratory of Quantum Theory and Applications of MoE, Lanzhou University, Lanzhou 730000, China}
\affiliation[a4]{MoE Frontiers Science Center for Rare Isotopes, Lanzhou University, Lanzhou 730000, China}
\affiliation[a5]{Research Center for Hadron and CSR Physics, Lanzhou University and Institute of Modern Physics of CAS, Lanzhou 730000, China}

\begin{abstract}
Recent CMS observations in the di-$J/\psi$ invariant mass spectrum reveal new structures at 6638 MeV ($X(6600)$)) and 7134 MeV, and confirm the $X(6900)$ at 6847 MeV. These findings provide crucial insights into four-charm-quark matter. The CMS Collaboration's result serves as a testament to the progress in exploring exotic hadronic matter.
\end{abstract}



\end{frontmatter}

Twenty years have passed since the observation of the $X(3872)$ as the first $XYZ$ charmonium-like state. With the accumulation of experimental data, more and more new hadronic states have been discovered, inspiring extensive attention to exotic hadronic matter, including multiquarks, hybrid, and glueballs \cite{Chen:2016qju,Chen:2022asf}. Taking this opportunity, it is an appropriate time to build the “Particle Zoo 2.0” version, which is the main task of the current study of hadron spectroscopy. More importantly, these joint experimental and theoretical efforts will further advance our understanding of the non-perturbative behavior of the strong interaction by these joint. In fact, representing the precision frontier of particle physics, hadron spectroscopy is entering the high-precision era.

\begin{figure}[htbp]
\centering
 \includegraphics[width=8.2cm]{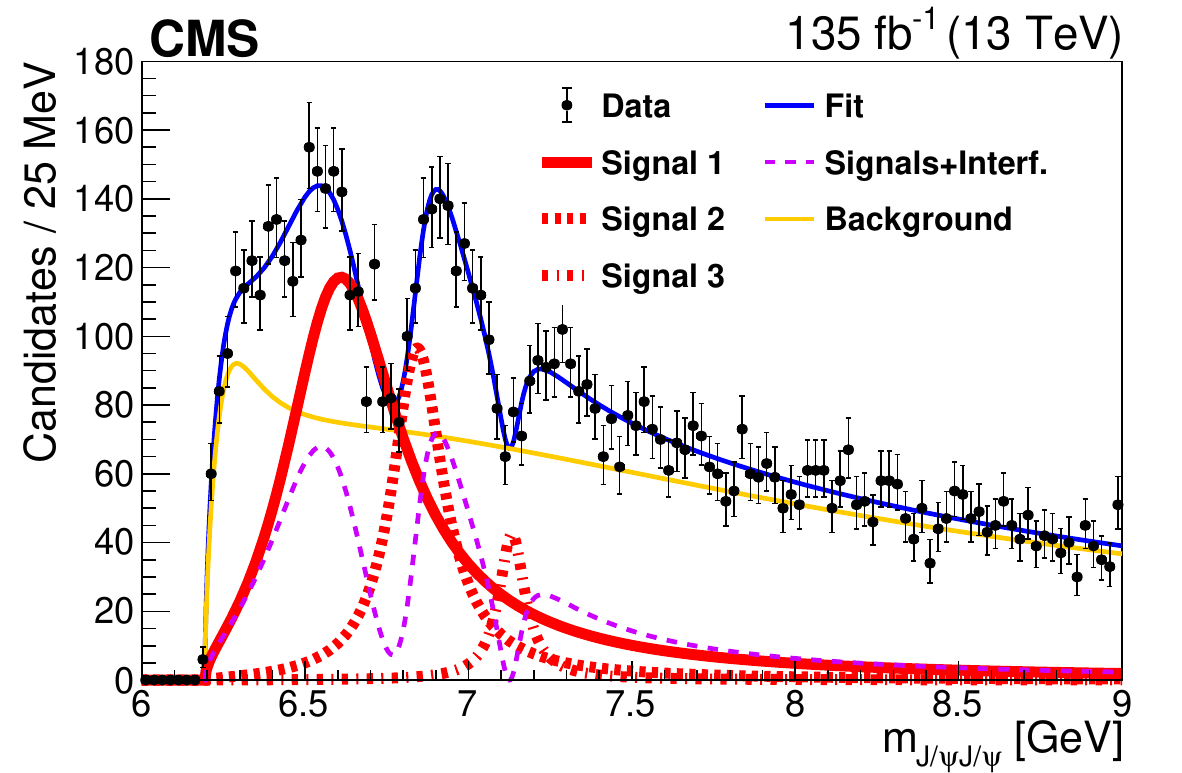}\\
\includegraphics[width=8.2cm]{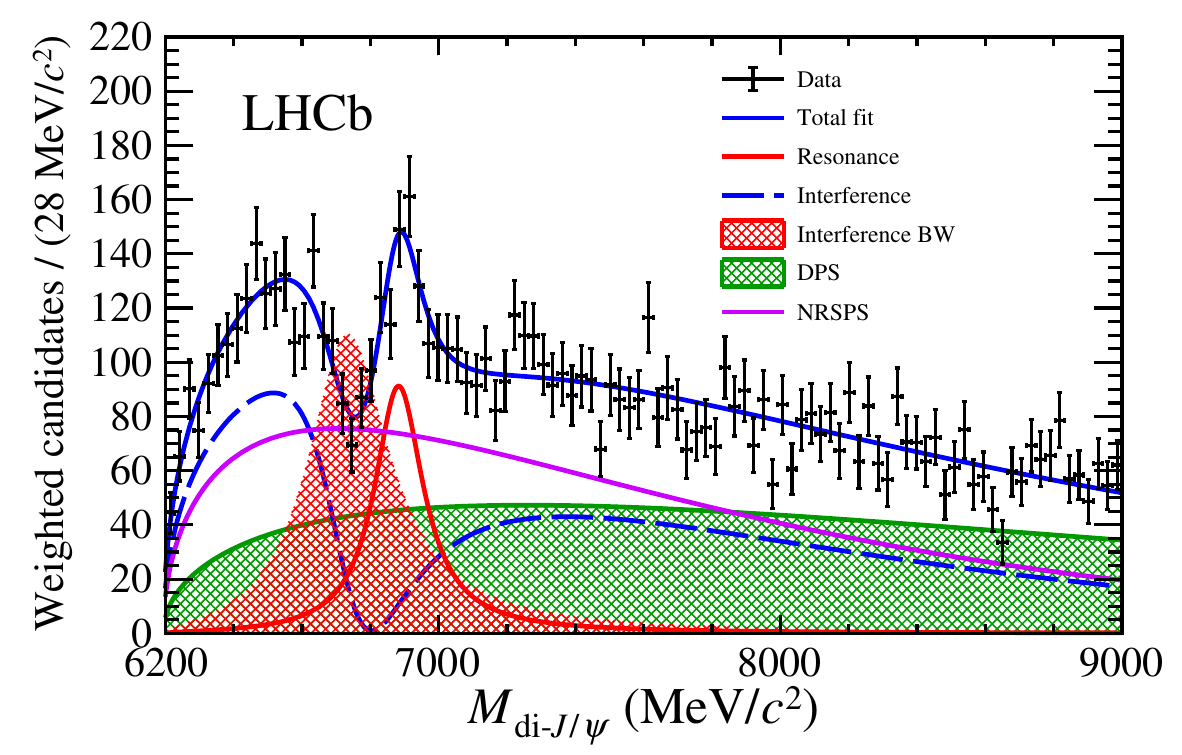}\\
\includegraphics[width=7.7cm]{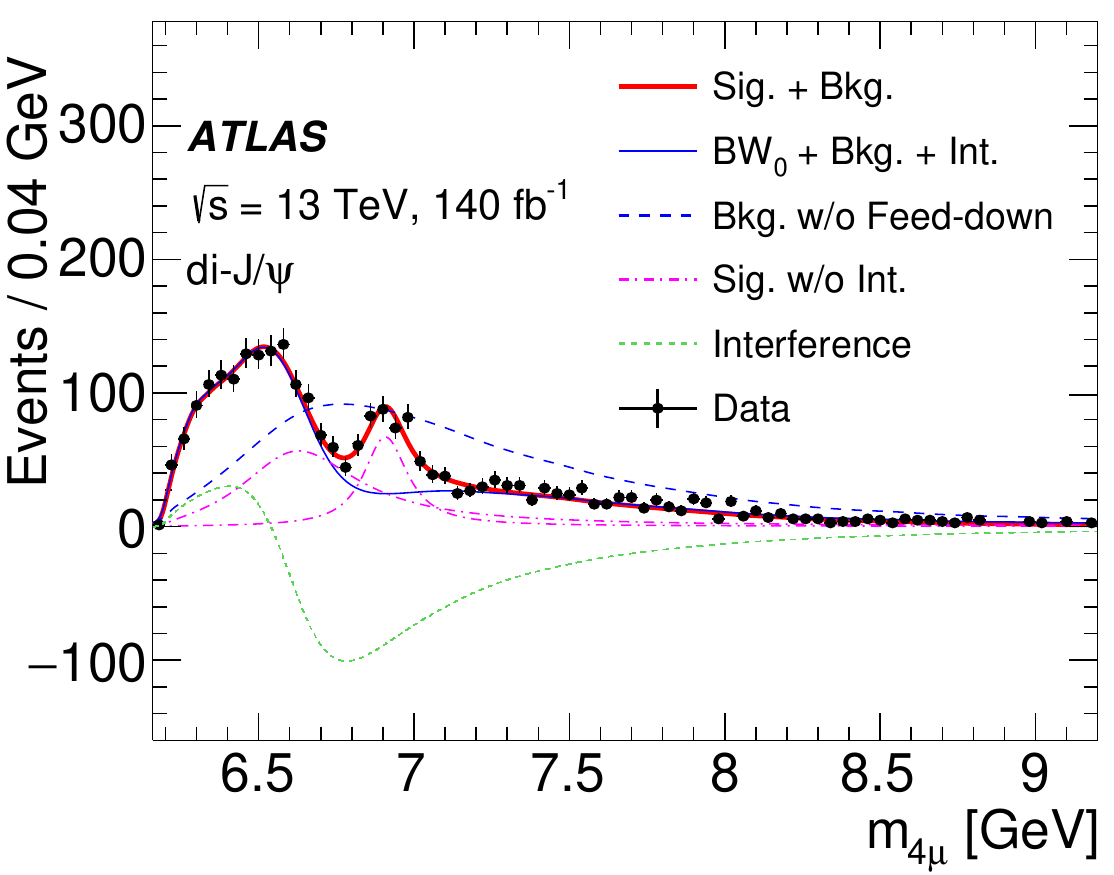}\\
\caption{A comparison of CMS \cite{CMS:2023owd}, LHC \cite{LHCb:2020bwg}, and ATLAS \cite{ATLAS:2023bft} results for the di-$J/\psi$ invariant mass spectrum. The CMS results show the fits consisting of three signal functions (BW1, BW2, and BW3) and one background component (NRSPS+DPS+BW0) \cite{CMS:2023owd}. 
The fit that includes the interferences, where “Interfering BWs” refers to the total contribution of all the interfering amplitudes and their cross terms.
The upper plot gives the following result: $m[X(6600)]=6638 ^{+43+16}_{-38-31}$ MeV, $\Gamma[X(6600)]=440 ^{+230+110}_{-200-240}$ MeV, $m[X(6900)]=6847^{+44+48}_{-28-20}$ MeV, $\Gamma[X(6900)]=191^{+66+25}_{-49-17}$ MeV, and $m[X(7100)]=7134 ^{+48+41}_{-25-15}$ MeV, $\Gamma[X(7100)]=97 ^{+40+29}_{-29-26}$ MeV.
In the middle plot, if LHC assumes interference between the NRSPS continuum and the broad structure near the di-$J/\psi$ mass threshold, it gives results of $m[X(6900)]=6886 \pm 11 \pm 11$ MeV and $\Gamma[X(6900)]=168\pm 33\pm 69$ MeV \cite{LHCb:2020bwg}.
In the lower plot, the ATLAS results are shown \cite{ATLAS:2023bft}: purple dash-dotted lines represent the components of the individual resonances, while green short-dashed lines indicate the interferences among them. The results obtained are $m[X(6600)]=6.65 \pm 0.02 ^{+0.03}_{-0.02}$ GeV, $\Gamma[X(6600)]=0.44\pm 0.05 ^{+0.06}_{-0.05}$ GeV, and $m[X(6900)]=6.91 \pm 0.01\pm 0.01$ GeV, $\Gamma[X(6900)]=0.15\pm 0.03 \pm 0.03$ GeV \cite{ATLAS:2023bft}.}\label{CMS}
\end{figure}

A very recent observation of more enhancement structures in the di-$J/\psi$ invariant mass spectrum from $pp$ collisions, which is reported by the CMS Collaboration, is a typical example to illustrate the importance of precision~\cite{CMS:2023owd}. Compared to previous measurements of the di-$J/\psi$ invariant mass spectrum by the LHCb Collaboration~\cite{LHCb:2020bwg} and the ATLAS Collaboration~\cite{ATLAS:2023bft}, the CMS result reveals two new structures at 6638 MeV and 7134 MeV in the di-$J/\psi$ invariant mass spectrum (see Fig. \ref{CMS}), which is a result of a particular fit saturated by a sum of several Breit-Wigner distributions. The first enhancement, observed with a significance greater than $5 \sigma$, has been named as the $X(6600)$. Additionally, CMS reports a structure at 6847 MeV consistent with the $X(6900)$ previously observed by LHCb. 
These discoveries provide crucial insights into the formation of four-charm-quark matter, a unique form of matter in nature.

Prior to the CMS and LHCb observations, some pioneering theoretical work around the compact $cc\bar{c}\bar{c}$ tetraquarks was done using different approaches \cite{Chao:1980dv,Barnea:2006sd,Berezhnoy:2011xy,Chen:2016jxd}. However, reconciling the experimental result with most of the theoretical predictions of the compact fully charm tetraquark is still a challenge, since their mass discrepancy is obvious. In 
addition, there have been tentative attempts to study whether these observed enhancement structures as molecular-type fully charm tetraquarks with the charmonium components \cite{Lu:2023ccs}. 

It is found that the threshold positions of all allowed combinations of intermediate charmonium pairs in the di-$J/\psi$ energy region from 6.20 to 7.30 GeV can be assigned to four main regions, namely $(6.45-6.64)$, $6.783$, $(6.87-7.00)$ and $(7.03-7.13)$ GeV. It is interesting to note that these observed enhancements in the CMS data correspond exactly to the four characteristic energy regions mentioned above. In view of the above fact, a new coupled-channel dynamical mechanism based on the special reactions has been proposed, considering that all the possible combinations of a double charmonium produced directly by a proton-proton collision transition to the di-$J/\psi$ final state \cite{Wang:2020wrp,Dong:2020nwy}. 
The LHCb experimental data of the line shape of the di-$J/\psi$ invariant mass spectrum can be well described \cite{LHCb:2020bwg}. As indicated in Refs. \cite{Wang:2020wrp,Wang:2020tpt}, the search for fully-heavy structures in the di-$\Upsilon(1S)$, $J/\psi\psi(3686)$, $J/\psi\psi(3770)$, $\psi(3686)\psi(3686)$, and $J/\psi\Upsilon(1S)$ invariant mass spectra can effectively test the universality of this dynamical mechanism. Notably, the ATLAS Collaboration has found evidence of new enhancement structures in the $J/\psi\psi(3686)$ invariant mass spectrum \cite{ATLAS:2023bft}. With more precise data expected from LHC Run-3, this approach can be applied to replicate the distribution of the di-$J/\psi$ invariant mass spectrum \cite{Wang:2022jmb}. Consequently, we anticipate the first observation of fully-heavy structures in the $J/\psi\psi(3686)$ or similar invariant mass spectra by the ATLAS, CMS, and LHCb experiments. 

Further efforts are needed to decipher these fully charm structures in the di-$J/\psi$ invariant mass spectrum. There is challenge for experimental analysis when faced with high precision experimental data. As pointed out in the CMS article~\cite{CMS:2023owd}, although the model with interference is not the only solution for the two dips among the structures $X(6600)$, $X(6900)$, and $X(7100)$, a fitting interference model describes the CMS data well.
In the following, a precise measurement of the spin and parity would provide deeper insights into the nature of these structures. This high-precision experimental measurement by CMS requires a high-precision theoretical study. 
{More theoretical effort should be devoted to checking whether these are exotic hadronic states or the result of dynamical generation, as shown in previous tentative results \cite{Dong:2020nwy,Wang:2022jmb}, where not all visible structures in the measured line shape correspond to nearby poles in the amplitude that can be regarded as physical states.} 
We also expect further experimental results on fully heavy systems from CMS in the near future.

\section*{Acknowledgments}
This work is supported by  the National Natural Science Foundation of China under Grants No.~12335001 and No.~12247101, National Key Research and Development Program of China under Contract No.~2020YFA0406400, the 111 Project under Grant No.~B20063, the fundamental Research Funds for the Central Universities, and the project for top-notch innovative talents of Gansu province.

\end{document}